\documentclass[letterpaper]{article}
\usepackage{aaai}
\usepackage{times}
\usepackage{helvet}
\usepackage{courier}
\usepackage{graphicx}
\usepackage{wrapfig}
\usepackage{multirow}

\usepackage{graphicx}
\usepackage{algorithmic}
\usepackage{algorithm}
\usepackage{amsmath}

\newcommand{\figref}[1]{Figure~\ref{#1}}

\newcommand{\be}{\begin{eqnarray} \begin{aligned}}
\newcommand{\ee}{\end{aligned} \end{eqnarray} }
\newcommand{\benn}{\begin{eqnarray*} \begin{aligned}}
\newcommand{\eenn}{\end{aligned} \end{eqnarray*} }
\usepackage{color}

\long\def\remove#1{}

\title{What Stops Social Epidemics?}
\author{Greg Ver Steeg, Rumi Ghosh \and Kristina Lerman \\
Information Sciences Institute \\
University of Southern California \\
{\{gregv,ghosh,lerman\}@isi.edu},
}

\begin{document}
\maketitle

\begin{abstract}

Theoretical progress in understanding the dynamics of spreading processes on graphs suggests the existence of an epidemic threshold below which no epidemics form and above which epidemics spread to a significant fraction of the graph.
We have observed information cascades on the social media site Digg that spread fast enough for one initial spreader to infect hundreds of people, yet end up affecting only 0.1\% of the entire network.
We find that two effects, previously studied in isolation, combine cooperatively to drastically limit the final size of cascades on Digg.
First, because of the highly clustered structure of the Digg network, most people who are aware of a story have been exposed to it via multiple friends.  This structure lowers the epidemic threshold while moderately slowing the overall growth of cascades.
In addition, we find that the mechanism for social contagion on Digg points to a fundamental difference between information spread and other contagion processes: despite multiple opportunities for infection within a social group, people are less likely to become spreaders of information with repeated exposure. 
The consequences of this mechanism become more pronounced for more clustered graphs. Ultimately, this effect severely curtails the size of social epidemics on Digg.
\end{abstract}

\section{Introduction}
Many diverse phenomena can be modeled as contact processes, including adoption of new ideas~\cite{Rogers03,Bettencourt05},  spread of infectious disease~\cite{Anderson91,Hethcote00} and behaviors~\cite{Christakis07obesity,Christakis08smoking}, computer virus epidemics on the Internet~\cite{Castellano10}, word-of-mouth recommendations~\cite{Goldenberg01}, viral marketing campaigns~\cite{Kempe03,Iribarren09}, and information cascades in online social networks~\cite{Lerman10icwsm}. A contact process is simply a diffusion of activation on a graph, where each activated, or ``infected,'' node can infect its neighbors with some probability given by the $transmissibility$. Given their prevalence, contact processes and the effect of network topology on their dynamics have been widely studied. %
One of the more important results is the existence of an \emph{epidemic threshold}~\cite{WangFaloutsos,PrakashFaloutsos,Chakrabarti08,Castellano10},  i.e., the critical value of transmissibility above which
large number of nodes in the graph are eventually infected.
In graphs with power-law degree distribution, a common property of social networks~\cite{Amaral00}, large degree heterogeneity speeds up epidemics~\cite{VespignaniBook,Lloyd-Smith05}, resulting in the  vanishing epidemic threshold in the limit of very large graphs~\cite{satorras2001}.  This result has alarming implications for propagation of viruses in human populations and computer networks: any outbreak, even one that is not very virulent, will spread to infect large number of nodes.

Until recently, obtaining empirical data to study contact processes involved laborious surveys~\cite{ValenteBook} and contact traces~\cite{Lloyd-Smith05,Christakis07obesity,Christakis08smoking}, which made analysis of their statistical properties impractical.
The proliferation of online social networks on sites such as Facebook, Twitter, and Digg, where users explicitly declare social links and actively  spread information, gives us a unique opportunity to quantitatively study dynamics of contact processes. We collected data from the social news aggregator Digg detailing how interest in more than 3,500 stories spreads through Digg's social network~\cite{Lerman10icwsm}. A user becomes $infected$ by $digging$ (i.e., voting for) a story and exposes her network neighbors to it. Each neighbor may in turn become infected (i.e., vote),  exposing her own neighbors to it, and so on. This way interest in a story cascades through Digg's network.

This data enables us to trace the flow of information along social links and quantitatively study dynamics of information spread on a network.
We find that the vast majority of cascades grow far slower than expected from their initial spread and fail to reach ``epidemic'' proportions.
To understand why, we simulate information cascades on the Digg graph and on a synthetic graph constructed to have similar properties. We compare results to theoretical predictions and properties of real cascades on Digg. We find that while network structure somewhat limits the growth of cascades, a far more dramatic effect comes from the social contagion mechanism. Unlike the standard cascade models used in previous works on the spread of epidemics, repeated exposure to the same story on Digg does not make the user more likely to vote for it. {Furthermore, this effect becomes quite significant due to the structure of the Digg graph which results in repeated exposure for most users.} 
{While the effect of clusters on epidemics has been studied, and alternate contagion mechanisms on social networks have been observed~\cite{www11-hashtags}, no one has studied their interaction and noticed that they constructively interfere to drastically limit epidemic size.}
We define an alternative cascade model that fits empirical observations and show that in simulation it reproduces the observed properties of real information cascades on Digg. 
\section{Information Cascades on Digg}
\label{sec:digg}
The social news aggregator Digg (http://digg.com) is one of the oldest and more popular social media sites.  Digg allows users to submit links to  news stories and vote for, or \emph{digg}, them. There are multiple submissions every minute, many thousands a day.  A newly submitted story goes to the {upcoming} stories list, where it remains for 24 hours, or until it is promoted to the {front page}. Digg selects about a hundred of these stories every day to feature on its front page. Although the exact promotion mechanism is secret, it appears to take into account the number and the rate at which story receives votes.

Digg also allows users to designate friends and track their activities. The \emph{friends interface} shows  stories a user's friends recently submitted or voted for. The friendship graph is directed. When user $j$ lists user $i$ as a \emph{friend}, $j$ can watch the activities of $i$ but not vice versa. We call $j$ the \emph{fan}, or the follower, of $i$. A newly submitted story is visible in the upcoming stories list, as well as to submitter's fans through the friends interface. With each vote it also becomes visible to voter's fans.
{In the event that $j$ has $n$ friends who have voted for a story, the story appears in their interface with a colored badge with the number $n$ emblazoned on it.}
We used Digg API to collect data about 3,553 stories promoted to the front page in June 2009.\footnote{The data set is available at\\http://www.isi.edu/\textasciitilde lerman/downloads/digg2009.html} 
The data associated with each story contains story title, story id, link,  submitter's name, submission time, list of voters and the time of each vote, the time the story was promoted to the front page. In addition, we collected the list of voters' friends.

We define an \emph{active user} as any user who voted for at least one story on Digg during the data collection period. There were 139,409 active users, of which 71,367 designated at least one other user as a friend. We extracted the friends of these users and reconstructed the fan network of active users, i.e., a directed graph of active users who are watching activities of other users. There were 279,634 \remove{258,218 from anonymous data} nodes in the fan network, with 1,731,658 links.

\begin{figure}[htb] %
   \centering
   \begin{tabular}{c} %
   \includegraphics[width=0.75\columnwidth]{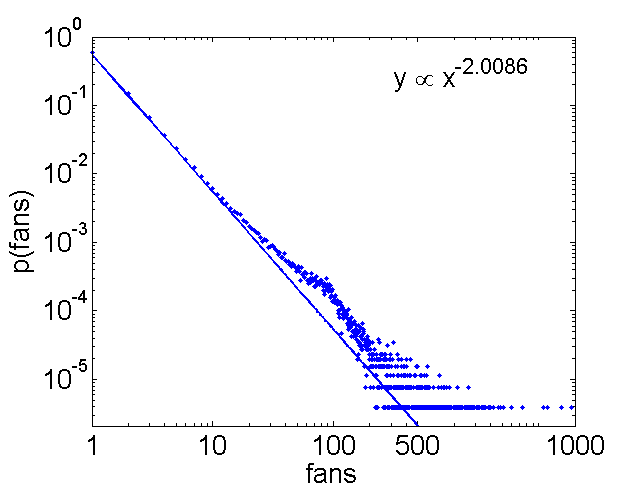} \\
   (a) \\
   \includegraphics[width=0.75\columnwidth]{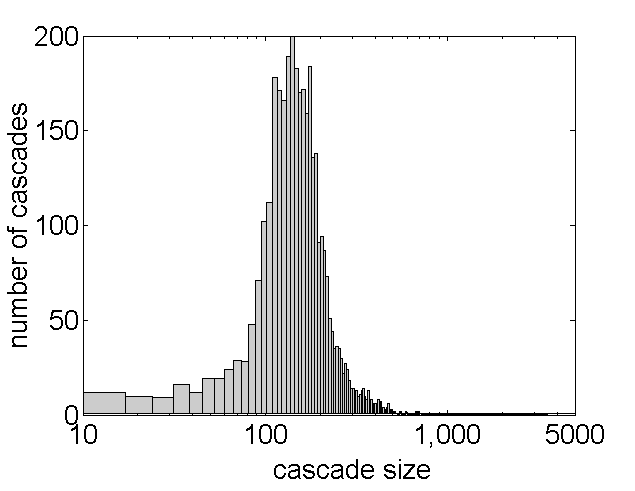} \\
  (b)
   \end{tabular}
   \caption{Properties of Digg: (a) Degree distribution of the fan network, (b) distribution of principal cascade size.}
   \label{fig:degreedist}
\end{figure}

\subsection{Degree Distribution}
Figure~\ref{fig:degreedist}(a) shows the distribution of the number of active fans per user. This distribution has a scale-free shape that is common to degree distributions of social and other complex networks~\cite{Amaral00,Clauset09} and is well described by a power law of the form $p(k) \propto k^{-\gamma}$ with $\gamma \approx 2$.

\subsection{Cascade Size}
The spread of a story through Digg's social network can be described as a social contagion process where votes spread between friends on the network.
As interest in a story spreads, it may generate many cascades from independent seeds.
For each story, using the methodology proposed in~\cite{Ghosh11wsdm}, we extracted the cascade that starts with the submitter and includes all voters who are connected to the submitter either directly or indirectly via the fans network. We call this the \emph{principal cascade} of the story. Figure~\ref{fig:degreedist}(b) shows the  distribution of principal cascade sizes of stories in our sample. This distribution is well described by a log-normal function with the mean of 156. Note that most of the cascades are smaller than 500, and only three are bigger than 1,000.

A story will typically generate multiple, even hundreds of, cascades~\cite{Ghosh11wsdm}. Aggregating over all cascades gives the total story popularity. This quantity is well fit by a log-normal with the mean 614. Only 15 stories in our sample of 3,553 received more than 6,000 votes, and only one more than 9,000. The most popular story in our sample, and one that generated the biggest principal cascade, was about Michael Jackson's death. It received more than 24,000 votes.
The distribution of story popularity is remarkably similar to that obtained by Wu and Huberman~\cite{Wu07} from a sample of over 30,000 stories promoted to Digg front page in 2006. In their sample, the most common value of story popularity was around 500 votes (with a maximum around 4,000 votes).

\subsection{What Limits Cascades on Digg?}
The observations above present a puzzle: why are information cascades on Digg so small? In our sample, only one cascade, about Michael Jackson's death, can be said to have reached epidemic proportions, i.e., reaching a significant fraction of active Digg users (in this case, about 5\%). The majority of the cascades for the remaining stories reached fewer than 0.1\% of active Digg users. On the other hand, the cascades did spread fast enough to infect hundreds of users.
{This disparity becomes more striking in the next section where we show that typical epidemic models predict that stories will reach an order of magnitude more voters than we observe on Digg.}

There are a number of factors that could explain why information cascades on Digg are so small.
Perhaps Digg users modulate transmissibility of stories and keep them small to prevent information overload.
On the other hand, transmissibility could diminish in time, either because of novelty decay~\cite{Wu07} or decrease in visibility of stories as new stories are submitted to Digg~\cite{Hogg09icwsm}.
Perhaps the structure of the network (e.g., clustering or communities) limits the spread of information.
Or it could be that the mechanism of social contagion, i.e., how people decide to vote for a story once their friends vote for it, prevents stories from growing on Digg. In addition, users are active at different times, and heterogeneity of their activity could be another explanation.

In this paper we examine some of these alternate hypotheses through simulations of contact processes on networks and empirical study of real cascades on Digg. {Ultimately, we are able to identify the factors that allow us to closely reproduce the observed behavior on Digg.}

\section{Analysis of Simulated and Real Cascades}

We now proceed to describe two effects that interact synergistically to severely limit the size of cascades on Digg.
First, due to the highly clustered structure of the Digg graph, most nodes end up being exposed to a story multiple times, even while the story fails to propagate outside of a cluster.
Second, we observe that, contrary to many contagion models, repeated exposure to a story does not make a user more likely to vote on it.
We compare observed Digg cascades to simulations and theoretical results for
standard models of spreading processes on graphs, highlighting the profound impact these effects have on the final size of cascades.

\subsection{Network Structure}\label{sec:structure}

\begin{figure}[htbp] %
   \centering
   \includegraphics[width=0.75\columnwidth]{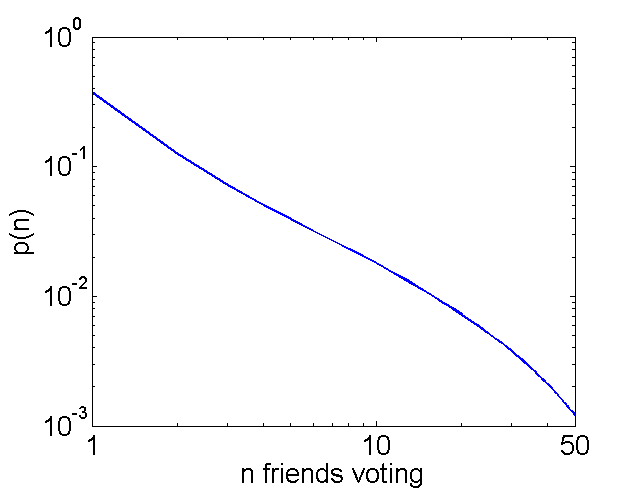}
   \caption{For nodes who were exposed to a story, the average number of friends who voted on the story.
}
   \label{fig:distfriendsvoting}
\end{figure}

A traditional measure of graph clustering like the (Watts-Strogatz) clustering coefficient, which is based on  the number of triangles in a graph,  yields an unremarkable $0.0924$ for the Digg graph.  In practice, we find clustering effects to be far more pronounced than this measure suggests,
potentially reflecting high variance of the clustering coefficient across all nodes.
For every node that sees a story from one of its friends, we count the total number of that node's friends who voted on the story (or, if the node itself voted, the total number of friends who voted on the story before it did). The distribution of this quantity in \figref{fig:distfriendsvoting} shows that a solid majority of $ \sim63 \%$ of exposed users have more than one friend voting on a story, with some having dozens of infected friends. This is especially remarkable when one considers theoretical results that model social contagion as branching processes, e.g., a Galton-Watson process \cite{Iribarren09}. That model assumes cascades spread in a tree-like fashion, so that each node has only one friend voting before it.

\subsection{Simulations}\label{sec:simulation} To gauge the effect of graph structure on the spread of cascades, we construct a graph with the same number of nodes and require that each node has the same degree as its counterpart in the Digg graph. We used the directed configuration model from \cite{configurationmodel} to create a random graph with a given degree sequence. This method preserves the degree distribution of the original Digg graph while destroying degree correlations and cluster structure in the graph. By simulating cascade processes on the original and randomized network, we can measure the effect graph structure has on cascade size.

We begin with an independent cascade model widely used to study diffusion processes on networks~\cite{satorras2001,Kempe03,Iribarren09}.
For later comparison we point out similarities to the susceptible-infected-removed (SIR) model in the epidemic literature~\cite{Hethcote00,VespignaniBook}.
We start with a single seed node who has voted for a story. By analogy with epidemic processes, we call this node \emph{infected}. The \emph{susceptible} fans of the seed node decide to vote on the story with some probability given by the \emph{transmissibility} $\lambda$. Since every node can only vote for the story once, at this point, the seed node is removed, and we repeat the process with the newly infected nodes. Note that a node who is a fan of $n$ voting nodes, has $n$ independent chances to become infected, but in this model once a node votes on a story, it only has one chance to spread it to its fans before it is removed.\footnote{In the epidemic literature this is equivalent to setting the recovery rate $\mu = 1$ and the infection rate $\beta = \lambda$.} Intuitively, this assumption implies that you are more likely to vote on a story if many of your friends vote on it.

Starting with random seed node, we generated 100,000 cascades using a transmissibility picked uniformly from the range [0, 0.01] and 40,000 cascades with transmissibility picked uniformly from the range [0.01, 0.03]. Each time a node is infected, it will infect each of its fans independently with probability $\lambda$. Additionally, we model the time between seeing a story and voting for it as a random variable pulled uniformly from some interval.

\begin{figure}[htbp] %
   \centering
   \includegraphics[width=0.75\columnwidth]{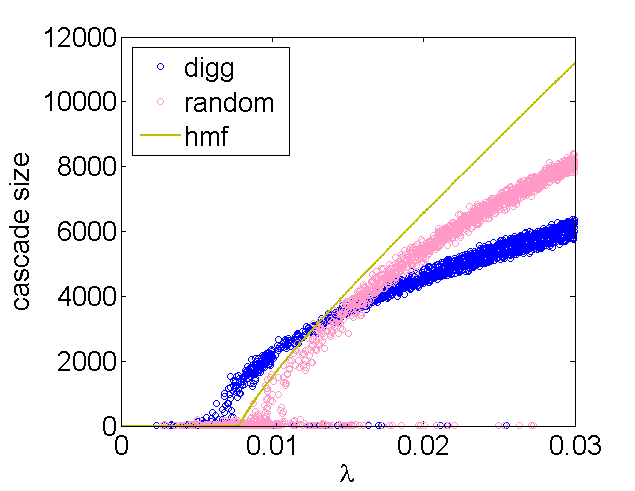}
   \caption{Cascade size as a function of transmissibility $\lambda$ for simulated cascades on the Digg graph and the randomized graph with the same degree distribution (see section on simulations). Heterogeneous mean field predicts cascade size as a fraction of the nodes affected. The line (hmf) reports these predictions multiplied by the total number of nodes in the Digg network. }
   \label{fig:cascadesize}
\end{figure}

After some time, no new nodes are infected, and the cascade stops. Because the graph is finite, the cascade is guaranteed to stop eventually. The final number of infected nodes gives cascade size. These are shown in \figref{fig:cascadesize}, where each point represents a single cascade with the $y$-axis giving the final cascade size and the $x$-axis denotes the transmissibility $\lambda$. We only keep  cascades with more than 10 infected nodes (votes).
Blue dots represent cascades on the original Digg graph while pink dots represent cascades on a randomized version of the Digg graph. In both simulations, there exists a critical value of $\lambda$, the \emph{epidemic threshold}, below which cascades quickly die out and above which they spread to a significant fraction of the graph. Note that even above epidemic threshold, cascades that start in an isolated region of the graph will die out.

\subsection{Theoretical results}\label{sec:theory}
The location of the epidemic threshold is accurately calculated for both the Digg and randomized graph using the inverse of the largest eigenvalue of the adjacency matrix of the graph~\cite{WangFaloutsos}. For the original graph this gives $\lambda_c^{digg} = 0.00587$, while for the randomized graph this gives $\lambda_c^{rand} = 0.00928$.

As we noted previously, this process should be accurately modeled by the SIR model of epidemics. In the limit of large graphs, if we assume that a node's behavior is defined by its degree (with no degree correlations), we can calculate the expected size of cascades using heterogeneous mean field (HMF) theory~\cite{moreno}. Based on observations of the Digg network shown in \figref{fig:degreedist}(a), we pick a degree distribution $p(k) \propto k^{-2}$, with a cut-off on the maximum degree, $k_{max} = 10^3$. This prediction is depicted with the gold line in \figref{fig:cascadesize}.  Both the threshold and growth accurately characterize the randomized graph. Note that HMF applies in the large graph limit. Because the randomized graph is still finite, some clustering inevitably occurs (it has a clustering coefficient of about $0.02$), decreasing the cascade size from the HMF prediction.

\subsection{Comparison of theory and simulation}
If one assumes Digg's graph structure consists of dense clusters, the effects on cascades in the independent cascade model are quite intuitive. It is easier for a story to take off within a smaller, more tightly connected community, thereby lowering the epidemic threshold. This also explains why the majority of people exposed to story are exposed to it from multiple sources. On the other hand, for cascades to grow very large it is better to have a more homogeneous link structure to reach all parts of the graph quickly. Ultimately, clusters have the effect of marginally decreasing the size of cascades by sequestering an infection in one part of the graph.

Comparing the theoretical and simulation results for cascades in \figref{fig:cascadesize} to the observed distribution of cascade sizes in \figref{fig:degreedist}(b) highlights the aforementioned puzzle. Why are cascades so small? According to our cascade model, only transmissibilities in a very narrow range near the threshold produce cascades of the appropriate size of $\sim 500$ votes. Is there some sort of critical behavior that tunes transmissibilities to be exactly near the threshold? Our subsequent analysis suggests a more intuitive possibility.

\subsection{Friend saturation model for Digg}

\begin{figure}[htbp] %
   \centering
   \includegraphics[width=0.75\columnwidth]{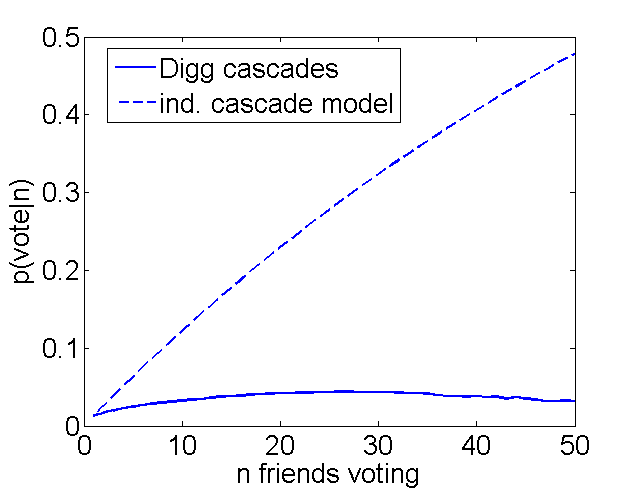}
   \caption{Characteristics of voting on Digg.
Probability a user votes given $n$ friends voted given by the independent cascade model and actual voting behavior on Digg (averaged over all cascades).
}
   \label{fig:friendsvoting}
\end{figure}

As we previously noted, modeling the spreading process as a branching process assumes that each node has only one voting friend. In that case, the definition of transmissibility, $\lambda$, is unambiguous: the probability of voting for a story if your friend voted. As we saw from \figref{fig:friendsvoting}, most people exposed to a story are exposed multiple times. In that case, even if we maintain the definition of transmissibility in the case of one friend voting, there is some freedom to model the effect of having multiple voting friends. The most straightforward generalization, which we used in the last section, is the independent cascade model(ICM) which says that if a node has $n$ voting friends then it has $n$ independent chances to vote. Therefore,
$$p_{ICM}( \mbox{vote} | n \mbox{ friends voted}) = 1-(1-\lambda)^n.$$

We can measure this quantity on Digg. To do so, we considered all cascades with more than 10 votes. We isolated the users in a cascade who had exactly $n$ friends voting and did not vote versus people with $n$ friends voting on the story before they themselves voted. For a given $n$, the percentage of people voting is depicted with a solid line in \figref{fig:friendsvoting}. For $n=1$, the percentage of users voting was $1.3\%$, suggesting an average transmissibility of $\lambda = 0.013$. The dashed line depicts $p_{ICM}( \mbox{vote} | n \mbox{ friends voted})$ for this value of transmissibility. Even for $n=2$, ICM overestimates the probability of a vote, and by $n=10$, a relatively common occurrence, ICM is an order of magnitude too large.

Clearly, \figref{fig:friendsvoting} shows that multiple exposures to a story only marginally increase the probability of voting for it.
The effect of multiple recommendations quickly saturates
 and would be better approximated as constant $$p_{FSM}( \mbox{vote} | n \mbox{ friends voted}) = \lambda,$$
 for $n\geq1$. We will refer to this simplified model for generating cascades as the ``friend saturation model'' (FSM).
We point out that $p( \mbox{vote} | n \mbox{ friends voted})$ actually contains two factors: the probability that you visit Digg and see that your friend(s) voted on a story, and the probability you vote on the story given that you did visit.
In fact, a careful examination of \figref{fig:friendsvoting} suggests that a more sophisticated model of behavior might include some small marginal increase in voting probability from multiple voting friends, balanced by a marginal decrease from having many friends.  We stick to the simpler model for simplicity, and because our fundamental result is not sensitive to these details.

\subsubsection{Simulation of FSM}
We can repeat the simulation procedure of the previous section. This time, though, after a node is exposed to a story from one of its friends (voting with probability $\lambda$), if the node chooses not to vote, it will not vote in the future even if it is exposed to the story again.
We generated 100,000 cascades with transmissibility picked uniformly from the range [0, 0.04]. Again, we only keep cascades with more than 10 votes.

\begin{figure}[htbp] %
   \centering
   \includegraphics[width=0.75\columnwidth]{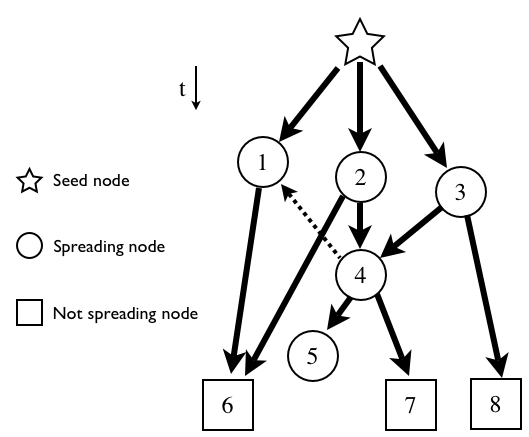}
   \caption{An example of a cascade for some story. An arrow from node $i \rightarrow j$ means that $j$ is a fan of $i$, and hence sees any stories that $i$ votes on. Time decreases down the y-axis so that the y position of a circular node indicates the time at which the node voted for this story. The dotted line indicates that although 1 is a fan of 4, this has no effect because 1 has already voted for this story before 4 spreads it to 1 again. The cascade terminates in nodes that do not vote for the story or nodes that vote but do not have any fans.}
   \label{fig:cascade}
\end{figure}

\subsubsection{Inferring Transmissibility of a Cascade}
Assuming that cascades on Digg spread according to the FSM process, we can infer the transmissibility of actual cascades.
Referring to Fig.~\ref{fig:cascade}, we label the nodes in a cascade in order of voting $i=1,\ldots,v$, where there are $v$ total nodes who vote on the story (not counting the seed node), and $i = v+1,\ldots,w$ label the $w-v$ nodes (watching) who are exposed to the story but do not vote on it.

\begin{figure}[htbp] %
   \centering
   \includegraphics[width=0.75\columnwidth]{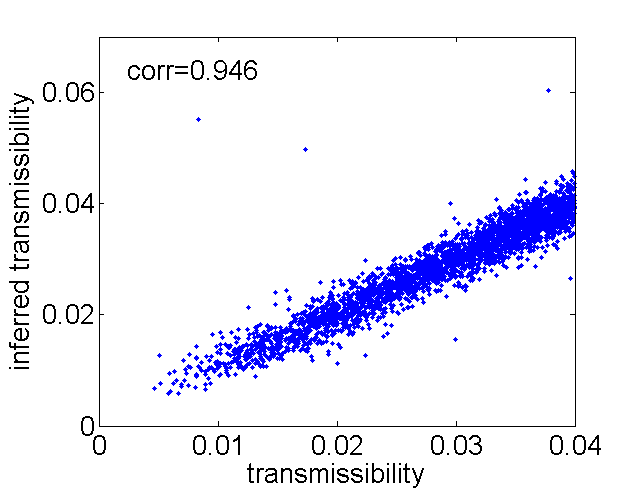}
   \caption{ Inferred versus actual transmissibility for simulated cascades in the FSM.}
   \label{fig:lamvsinf}
\end{figure}

According to the FSM, each node votes for a story that is spread to it with probability $\lambda$, and we can read the probability of a cascade directly from the graph.
\be
p(\mbox{cascade} | \lambda) &=& (1-\lambda)^{w-v} \lambda^v
\ee
Given a cascade, the maximum likelihood value for $\lambda$ is
\begin{equation}
\lambda_{inf} = \mbox{argmax}_\lambda p(\lambda | \mbox{cascade}) = \frac{v}{w}.
\end{equation}
To test the accuracy of this inference, we compute the inferred values of $\lambda$ for simulated cascades and plot them against the actual values used in the simulation in \figref{fig:lamvsinf}. Pearson's correlation coefficient for these values is $0.946$.

\begin{figure}[htbp] %
   \centering
   \includegraphics[width=0.9\columnwidth]{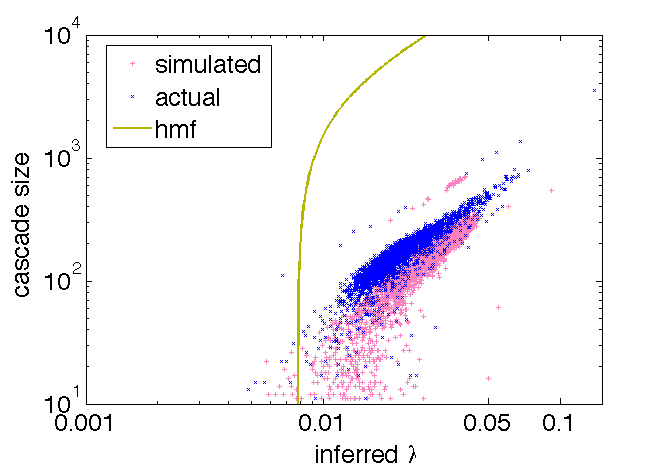}
   \caption{ Cascade size vs inferred transmissibility for simulated and real cascades on the Digg graph, this time plotted on a log-log scale to highlight the order of magnitude difference between these cascade sizes and predictions of the epidemic model (HMF, see text for details).}
   \label{fig:saturation-model}
\end{figure}

Pink dots in \figref{fig:saturation-model} plot size of simulated cascades generated according to the FSM versus inferred transmissibility.
Already, looking at the line plotting the HMF prediction from \figref{fig:cascade}, we see that cascade sizes are an order of magnitude smaller than for the independent cascade model.
Using this model, we can also infer transmissibilities for actual Digg cascades and we compare them on the same plot. The similarity is striking and the overlap so complete that most of the simulated cascade dots are covered.
Also, note that the threshold still lines up fairly well with HMF and eigenvalue prediction. At the beginning of a cascade, most people have not been exposed multiple times, so the FSM and independent cascade model differ very little, therefore we should not expect much change in the location of the threshold.

The inferred transmissibilities of actual Digg cascades are almost all above threshold. This is not surprising, given that we are analyzing stories that have been promoted to the front page and, therefore, have been found by Digg to be interesting to the community. Note that the largest cascade, one about Michael Jackson's death, also has the highest inferred transmissibility.

\subsection{Discussion}
\begin{figure}[htbp] %
   \centering
   \begin{tabular}{@{}c@{}c@{}}
   \includegraphics[width=0.5\columnwidth]{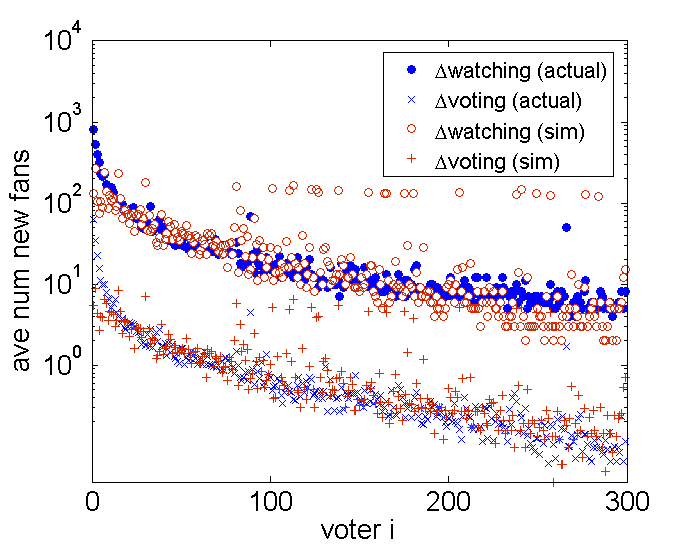} &
   \includegraphics[width=0.5\columnwidth]{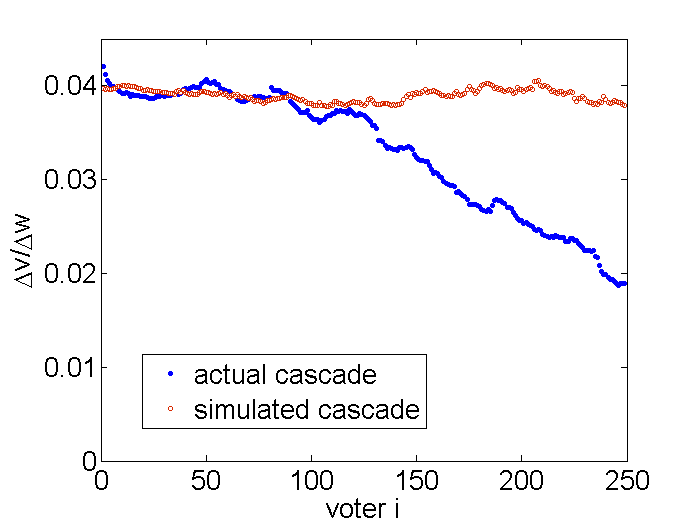} \\
   (a)&(b)
   \end{tabular}
   \caption{ Dynamics of transmissibility and fanout (a) Number of new fans who can see the story ($\Delta$watching) and who actually vote for the story ($\Delta$voting) vs time (voter $i$) for actual and simulated cascades. (b) Change in the estimated value of transmissibility for actual and simulated Digg cascades as a function of time. }
   \label{fig:dynamics}
\end{figure}

In epidemic models, population models, and other branching processes, the principal quantity of interest is the reproductive number, $R_0$. Intuitively, the reproductive number is just the average number of people infected by a single infected person. If $R_0 > 1$, each infection leads to another indefinitely, an epidemic. Whereas, if $R_0<1$, the infection will die out eventually.
Naively, the reproductive number should just be the average fanout, i.e., the average number of fans, times the transmissibility. For Digg, we have $\langle k \rangle\approx 6$ so $R_0 \approx 6 \lambda$. In that case, an epidemic threshold at $R_0=1 \rightarrow \lambda_c \approx 1/6$, much higher than we observe. It is well known, however, that heterogeneous degree distributions lower the threshold compared to this prediction~\cite{VespignaniBook}.

However, we can gain some intuition from this quantity if we view it as a dynamic quantity. FSM implies that the true fanout only includes the number of new fans (those that have not already been exposed to a story) and changes with time. \figref{fig:dynamics}(a) shows that with this definition the fanout is steadily decaying, both for actual and simulated cascades on the Digg graph. Effectively, this leads to a decrease in the reproductive number as well, so that a cascade that initially starts above the epidemic threshold may fall below it in time.

Additionally, in \figref{fig:dynamics}(b) we examine the dynamics of the transmissibility. We calculate the transmissibility for each voter by looking at the number of votes a story gets from their fans divided by the number of new fans the voter exposed the story to. We see this quantity is constant for simulated cascades as it should be by construction. For actual cascades, on the other hand, the transmissibility remains constant until about 100 people have voted, and then it begins to decline. This is another effect limiting the size of cascades. The decline could be due to decay of novelty~\cite{Wu07} or decrease in visibility~\cite{Hogg09icwsm} as a consequence of new stories being submitted to Digg.
Alternately, people may vote for stories mostly hoping to help them get ``promoted'' to the front page. After about 100 votes, a story is usually promoted, thereby offering less incentive to give it further votes.

Reproductive number is a product of fanout and transmissibility and on Digg both decrease with time. From this perspective, the slowdown of cascade growth is natural.

\section{Related Work}
Contact processes have been extensively studied in epidemiology, where compartmental models, such as SIS (susceptible-infectious-susceptible) and SIR (susceptible-infectious-recovered), have been used to model dynamics of epidemics. These models assume that everyone is in contact with everyone else in the population, and rate of  infection and recovery is uniform~\cite{Bailey:1975}. 
One approach to create more realistic models of interactions was to segregate population using different categorical features, such as age, sex and so on, and then treat the interactions within the subpopulations as homogenous and symmetric \cite{Hethcote1978}. Another approach to modifying homogenous models was to represent interactions between individuals as a directed graph \cite{Kephart91}, leading to a single mean field (MF) reaction equation. 
To relax the homogeneity assumptions further, and to take into account, the strong fluctuations in the connectivity distribution, the single MF equation is modified to a heterogenous mean field (HMF) reaction rate equation \cite{moreno}. In \cite{PrakashFaloutsos}, the authors conjecture  that for any virus propagation model (including SIS and SIR), the epidemic threshold depends only on the largest eigenvalue of the adjacency matrix of the network. However, \cite{Castellano10} argue that while this holds true for the SIS model, the HMF prediction in the SIR model seems to be much more accurate than the generic claim made in \cite{PrakashFaloutsos} for scale-rich networks. They  claim that on quenched scale-rich networks the threshold of generic epidemic models is vanishing or finite depending on the presence or absence of steady state. 

Another modified spreading process for social contagion that has been considered is the effect of adding ``stiflers''~\cite{VespignaniBook}. Similar to FSM, stiflers will not spread a story (rumor) no matter how many times they encounter it. Stiflers, however, are not merely desensitized to multiple exposures, they may actively convert spreaders or susceptible nodes into stiflers. This complicated dynamic can lead to drastic changes, e.g., the elimination of the epidemic threshold. In Digg, a fan who does not  vote on a story after multiple exposures, does not actively persuade the exposed and susceptible fans not to vote on a story. Hence, this model does not apply to the process of information diffusion on Digg.

The friend saturating model we have used to describe cascades on Digg is a special case of a broader class of models called ``decreasing cascade models''~\cite{Kempe03}. Several works have observed similar diminishing returns from friends in social networks. \cite{Leskovec07amazon} analyzed the usefulness of product recommendations on Amazon.com. They rarely found that anyone received more than a handful of recommendations for any product, and the marginal benefit of multiple recommendations, while product dependent, was typically sublinear (i.e. two recommendations did not make someone twice as likely to buy as one recommendation). Link formation was studied in \cite{Kossinets}, where they also found diminishing returns in the probability of befriending someone with whom one shares $n$ mutual friends, with saturation occurring around $n=5$. The probability of joining a group that $n$ friends have joined was studied in \cite{Backstrom06}, with saturation occurring for $n$ around  10-20.

\cite{Iribarren09} modeled viral email cascades using branching processes like the Galton-Watson  process and  the Bellman-Harris process. They argued that the topology  of the underlying social network is irrelevant in the prediction of cascade size.  This may hold true in the  tree-like cascades studied by the authors. However as stated previously, in Digg, dynamics of information propagation is not tree-like and  these models do not hold. Future work includes studying the impact of activity patterns on the information diffusion dynamics.

\section{Conclusion}
We have demonstrated a simple behavior model that strikingly reproduces the behavior of Digg cascades, while standard methods go awry.
Many network studies assume that graphs with locally tree-like behavior give a good approximation to real networks. In this case, we find that such methods wildly overestimate the size of cascades. If most of the people exposed to a story are exposed repeatedly, understanding how they are affected by repeat exposures is of paramount importance. On Digg, multiple exposures to a story are common and {have almost no effect on the probability of voting; this severely limits the size of epidemics.} %
Much remains to be studied: whether these results hold on other social networks, more sophisticated models of response to friends, the  time dependence of transmissibility, and more detailed analysis of the effect of graph structure on cascades.

\subsection*{Acknowledgements}
This work is supported in part by AFOSR under contracts 1295GNA276 and FA9550-10-1-0102, and by the NSF under grants 0915678 and 0968370.
We would also like to thank Tad Hogg for valuable input.

\section{Appendix}
We have recently~\cite{Ghosh11wsdm} proposed a framework to extract individual cascades of a spreading process. This approach uses a \emph{cascade generating function} to compute  a value for the spreading process at each time step that allows us to extract and reconstruct the structure of the cascade. We label nodes in the order they are activated. The {cascade generating function} is parameterized by the transmission rates $\alpha_{j,i}$  $ \forall j, i  \in [1,N]$ (where $N$ is the number of nodes involved in the spreading process), which  give the probability that a node $i$ activated at time ${t_i}$ will activate a connected node  $j$ at a later time $t_j$. For simplicity, we take $\alpha_{ji}=\alpha$.

Suppose there are $K$ seeds, which are activated at times $t=i_1,\ldots, i_K$. We set the value of the cascade generating function at the time a seed is activated to $\phi(i_p,\alpha)=1$.  If  node $i$ is activated at time $t_i$, the value of the cascade generating function at a later time when node $j$ is activated is:
 \begin{equation}
\phi(j, \alpha )=\sum_{i \in friend(j)}\alpha \phi(i, \alpha)=  \sum_{p=1}^{K}  f(j, i_p, \alpha) \phi(i_p, \alpha).
\label{eq:phi1}
\end{equation}
\noindent The function $f(j, i_p, \alpha)$ captures the cumulative effect the cascade generated at seed $i_p$ has on the node $j$ where $t_j> t_{i_p} $.   Note that the same node can belong to two or more cascades generated by different seeds, leading to the {collision of cascades} effect. All the nodes for which $f(j, i_p, \alpha) >0$ $\forall\ \alpha>0$, are the nodes that belong to the cascade generated by seed $i_p$.  Ref.~\cite{Ghosh11wsdm} describes an algorithm to extract a cascade generated by a seed that has $O(N)$ space and $O(dN)$ time complexity, where $d$ is the maximum degree and $N$ is the number of nodes in the network.

\end{document}